\documentclass[fleqn,10pt]{wlscirep}
\usepackage[utf8]{inputenc}
\usepackage[T1]{fontenc}
\title{Explainable Detection of Machine Generated Music and Early Systematic Evaluation}

\author[1,*,+]{Yupei Li}
\author[1,+]{Qiyang Sun}
\author[2]{Hanqian Li}
\author[1]{Lucia Specia}
\author[1,3]{Bj\"orn W. Schuller}

\affil[1]{Imperial College London, Computing, London, United Kingdom}
\affil[2]{Shandong University, Shandong, China}
\affil[3]{Technical Unitervisty of Munich, Munich, Germany}

\affil[*]{corresponding author yl7622@ic.ac.uk}

\affil[+]{these authors contributed equally to this work}

\begin{abstract}
Machine-generated music (MGM) has become a groundbreaking innovation with wide-ranging applications, such as music therapy, personalised editing, and creative inspiration within the music industry. However, the unregulated proliferation of MGM presents considerable challenges to the entertainment, education, and arts sectors by potentially undermining the value of high-quality human compositions. Consequently, MGM detection (MGMD) is crucial for preserving the integrity of these fields. Despite its significance, MGMD domain lacks comprehensive systematic evaluation results necessary to drive meaningful progress. To address this gap, we conduct experiments on existing large-scale datasets using a range of foundational models for audio processing, establishing systematic evaluation results tailored to the MGMD task. Our selection includes traditional machine learning models, deep neural networks, Transformer-based architectures, and State space models (SSM). Recognising the inherently multimodal nature of music, which integrates both melody and lyrics, we also explore fundamental multimodal models in our experiments. Beyond providing basic binary classification outcomes, we delve deeper into model behaviour using multiple explainable Artificial Intelligence (XAI) tools, offering insights into their decision-making processes. Our analysis reveals that ResNet18 performs the best according to in-domain and out-of-domain tests. By providing a comprehensive comparison of systematic evaluation results and their interpretability, we propose several directions to inspire future research to develop more robust and effective detection methods for MGM. We provide our codes and some samples on Github repository \url{https://github.com/myxp-lyp/Detecting-Machine-Generated-Music-with-Explainability-A-Challenge-and-Systematic-Evaluation}.
\end{abstract}
\begin{document}

\flushbottom
\maketitle
\thispagestyle{empty}

\section*{Introduction}
\label{sec:intro}

The rise of machine-generated music (MGM), driven by advancements in large language models (LLMs) and exemplified by platforms like MuseNet \cite{openai_musenet} and AIVA \cite{aiva2023}, is reshaping the music industry and cognitive science \cite{tan2025prompting}. While this technology offers transformative potential, it also raises critical concerns regarding originality, copyright, and the preservation of artistic value \cite{micalizzi2024artificial}. MGM enhances music production by stimulating creativity, providing compositional frameworks, and suggesting instrumental arrangements \cite{briot2020deep}, yet its capacity for rapid and large-scale output threatens traditional, artist-led creations that rely on time, effort, and individual expression. The growing dominance of Machine-generated styles risks homogenising the music landscape, reducing originality, and influencing public taste through overexposure to algorithmically driven patterns. To address these challenges, effective machine-generated music detection (MGMD) is essential for safeguarding the music community's creative diversity and fostering a balanced coexistence between human creativity and AI innovation.

Recent studies \cite{li2024audio} have critically examined the current landscape of MGMD research, emphasising the urgent need for further progress in this field \cite{cooke2024good}. %as highlighted by other works . , henry2024impacts}. 
The authors identify substantial gaps in the existing literature, particularly the lack of standardised systematic evaluation of various models, the absence of adequate comparative tools, and the challenges in establishing coherent research directions, especially in the context of explainable artificial intelligence (XAI). This led to the fact that the current MGMD tools are fragmented, often concentrated on narrow areas of investigation. For example, the Spectro-Temporal Tokens Transformer (SpecTTTra) \cite{rahman2024sonics} focuses solely on Mel-spectrum features to capture long-range dependencies. In contrast, Afchar \cite{afchar2024detecting} employs conventional convolutional methods, placing notable emphasis on the importance of out-of-domain testing and providing insights into output probability for improved interpretability. However, the interpretability remains insufficient, as the approach does not fully incorporate XAI principles. The authors of these innovative recent advancements have not released their models publicly, and their performance comparisons are limited and inadequately explained, highlighting the need for a comprehensive systematic evaluation of various models in this domain.

Recent studies \cite{li2024audio} suggest a potential transition from deepfake audio detection to MGMD, given the overlap in the audio features and their shared focus on binary classification. However, music presents unique challenges, particularly with lyrics. Several multimodal models have been developed for music generation, such as lyric-to-melody models \cite{Zhang_2022, ju2022telemelodylyrictomelodygenerationtemplatebased}, which show the efficacy of multimodal feature extraction. Despite these advancements, a comprehensive systematic evaluation of various models for multimodal models in this domain remains lacking.

To address the gaps identified above, we aim to assess the performance of widely used audio classification models as systematic evaluation, including traditional machine learning models, deep neural network models, Transformer-based models, and State space models (SSM) on the FakeMusicCaps dataset \cite{comanducci2024fakemusiccaps}. Subsequently, we conduct out-of-domain testing on the M6 dataset \cite{li2024m6multigeneratormultidomainmultilingual} and establish a systematic evaluation for multimodal models. Finally, we leverage and develop new XAI techniques to provide clearer insights into the models' performance.

Our contributions are:
\begin{itemize}
    \item To the best of our knowledge, this work presents an early and systematic evaluation of multiple models on the MGMD task. And we propose several future research directions. 
    \item We use multimodal models and out-of-domain testing to enhance the systematic evaluation's scalability and robustness.
    \item We propose a novel ensemble XAI strategy that aggregates consensus across multiple post-hoc attribution methods.
\end{itemize}

\section*{Related work}
\label{sec:related}
\subsection*{Fundamental detection models}
Recent studies have reviewed many %foundational models in music \cite{ma2024foundationmodelsmusicsurvey}, shedding light on the application of musicological features in traditional detection models. Building on these works, as well as research on 
deepfake audio detection models 
\cite{yi2023audiodeepfakedetectionsurvey, almutairi2022review}. %, li2022comparative}. 
Building on these and overlap application with MGMD, the following evaluation categories are proposed: traditional machine learning models, deep neural network models, Transformer-based models \cite{vaswani2017attention}, State space models (SSM), and multimodal models. The models we selected for evaluation are based on their common use of feature representations in music, particularly the Mel-spectrum.

\subsubsection*{Traditional machine learning models}
Traditional machine learning models are often employed as classifiers. While these models are frequently combined with additional feature processing techniques within a larger pipeline, they can also independently handle the entire task by utilising extracted features to predict labels directly. The \textbf{(Quadratic) Support Vector Machine ((Q)-SVM)} \cite{singh2021detection} is one of the most commonly utilised models, renowned for its relatively simple parameters. Empirical study \cite{agarwal2021deepfake}%, kharbat2019image} 
has demonstrated the effectiveness of QSVM for audio-based classification tasks.

\subsubsection*{Deep neural network models}
Convolutional Neural Networks (CNNs) \cite{lecun1995convolutional} have demonstrated capability in image-based feature analysis, leading to the development of models built upon them. \textbf{ResNet18} \cite{he2016deep}, by leveraging residual connections, addresses the vanishing gradient problem, with applications in deepfake detection shown in work \cite{rabhi2024audio}. Additionally, \textbf{VGG} \cite{simonyan2014very} employs deeper layers with small receptive fields, as applied to deepfake audio analysis \cite{mcuba2023effect}. To mitigate the inefficiency of such models, \textbf{SENet} \cite{hu2018squeeze} introduces squeeze-and-excitation blocks, which refine feature maps \cite{zhang2023improving}. Finally, \textbf{MobileNet} \cite{howard2017mobilenets} offers a more efficient alternative by utilising depthwise separable convolutions, with its application in synthetic audio detection \cite{wen2022multi}.

On top of models based on CNNs, music typically exhibits a long-form structure that incorporates melody and harmony. To better capture both temporal dependencies and spectral features simultaneously, a hybrid model combining \textbf{CNNs with LSTM} (Long Short-Term Memory) networks \cite{hochreiter1997long} has been proposed \cite{saikia2022hybrid} to enhance the model's ability to represent the sequential nature of musical data.

\subsubsection*{Transformer-based models}
Transformer-based models \cite{vaswani2017attention} have recently surpassed deep neural networks in feature extraction, owing to attention mechanism. The \textbf{Vision Transformer (ViT)} \cite{dosovitskiy2020image}, a variant of the Transformer architecture capable of processing images as input, has been successfully applied in audio deepfake detection \cite{ulutas2023deepfake}.

\subsubsection*{SSM}
SSMs are valued for their ability to model complex dynamic systems and capture temporal dependencies. \textbf{Mamba} \cite{gu2023mamba} employs linear-time sequence modeling to represent dynamic systems and capture long-range dependencies. \textbf{xLSTM} \cite{beck2024xlstmextendedlongshortterm} combines SSMs with extended LSTM and an attention-like mechanism to improve modeling of temporal patterns and contextual features. These design choices make both models well-suited for classification tasks.
\subsubsection*{Multimodal models}
Independent of the above choice, 
there exist multimodal models that focus on extracting audio (melody) and text (lyrics) features, albeit for other applications. Simonetta et al.\ \cite{simonetta2019multimodal} provide a comprehensive review of multimodal models in music information retrieval, highlighting the modality representation challenges. LLark \cite{gardner2024llarkmultimodalinstructionfollowinglanguage}, a multimodal LLM, has been proposed for instruction-following tasks; however, its text processing is designed for instructions rather than lyrics. %Similarly, Liu et al.\ \cite{liu2023m} propose multimodal models tailored for music generation. 
Despite architectural variations across applications, these models adhere to the principle of extracting features from each modality and subsequently fusing them. As for our systematic evaluation, the simplest approach is enough: features extracted using unimodal models are concatenated before the final classification step, a method known as the early fusion.
\subsection*{Explainable AI}
All the classifiers mentioned above make decisions based on their internal reasoning processes; however, without XAI, the decision-making process remains a black-box, limiting the transparency and interpretability of the results, which reduces their persuasiveness. 

Sun et al. \cite{sun2024explainableartificialintelligencemedical} recently provided a comprehensive overview of XAI techniques, focusing on their application in the medical domain. Prior studies have also explored the use of XAI in audio-based 
 deep learning \cite{10890370}, and acoustic feature analysis for audio deepfake detection \cite{bisogni2024acoustic}. However, XAI techniques specifically tailored to music feature understanding tasks, such as MGMD, remain underexplored. Nevertheless, insights from XAI applications in audio deepfake detection may prove valuable.

There are common foundational XAI techniques: \textbf{Occlusion Sensitivity} \cite{du2020learning} evaluates the importance of input regions by masking parts of the input and measuring the resulting changes in the model's output, but it is time-consuming. In contrast, \textbf{Integrated Gradients (IG)} \cite{sundararajan2017axiomatic} offer a more efficient way by accumulating gradients along a path from a baseline input to the actual input. With the help of gradients as well, \textbf{Gradient-weighted Class Activation Mapping (Grad-CAM)} \cite{selvaraju2017grad} leverages gradient information to localise critical regions in input images, highlighting areas relevant to the model’s decisions. The basic model of Grad-CAM, \textbf{CAM} \cite{zhou2016learning} directly utilises class-specific weights to identify discriminative regions, providing a class-focused perspective. However, CAM requires specific architectural modifications to the model. Unlike the model-specific XAI, model-agnostic model \textbf{LIME} \cite{ribeiro2016should} instead approximates local model behaviour using an interpretable surrogate model.

\section*{Experiments}
\label{sec:experiments}

\subsection*{Datasets}
%Datasets for MGMD tasks remain limited; however, we selected the widely used \textbf{FakeMusicCaps} dataset \cite{comanducci2024fakemusiccaps} for our evaluation. 
For basic evaluation, we selected \textbf{FakeMusicCaps} dataset \cite{comanducci2024fakemusiccaps}.
This dataset was designed to facilitate the detection of music authorship. It builds upon the MusicCaps dataset\footnote{https://paperswithcode.com/dataset/musiccaps}, from which human-annotated prompts were extracted and utilised to feed Text-to-Music models for generating MGM with five distinct music generation models. FakeMusicCaps contains 5,521 human-made samples and 27,605 MGM samples, each with a duration of 10 seconds.

For out-of-domain testing and multimodal evaluation, we selected the recently introduced \textbf{M6} dataset \cite{li2024m6multigeneratormultidomainmultilingual}, as FakeMusicCaps includes only background music without lyrics. The M6 dataset integrates existing human-made music and generates MGM using multiple generators in conjunction with commercial tools. This comprehensive dataset suits our needs. The f subset we use in M6 contains 1000 positive samples, 500 negative samples, with average duration 29.99/58.06 seconds respectively. 

We split the dataset into training, validation, and test sets for each type, with proportions of 0.8×0.8, 0.8×0.2, and 0.2, respectively. To facilitate further analysis, all audio files in WAV format were processed with the \emph{librosa} \cite{mcfee2015librosa} library to extract the Mel spectrogram representations. To ensure fair experimental comparisons, the same data split was maintained across all experiments. Additionally, all audio samples were resampled to a uniform rate of 16,000\,Hz.

\subsection*{Quantitative results}
We perform quantitative results to evaluate all models, with detailed hyperparameter choices in supplementary materials.
\subsubsection*{In-domain test}
\begin{figure}[h]
    \centering
    \includegraphics[width=0.8\linewidth]{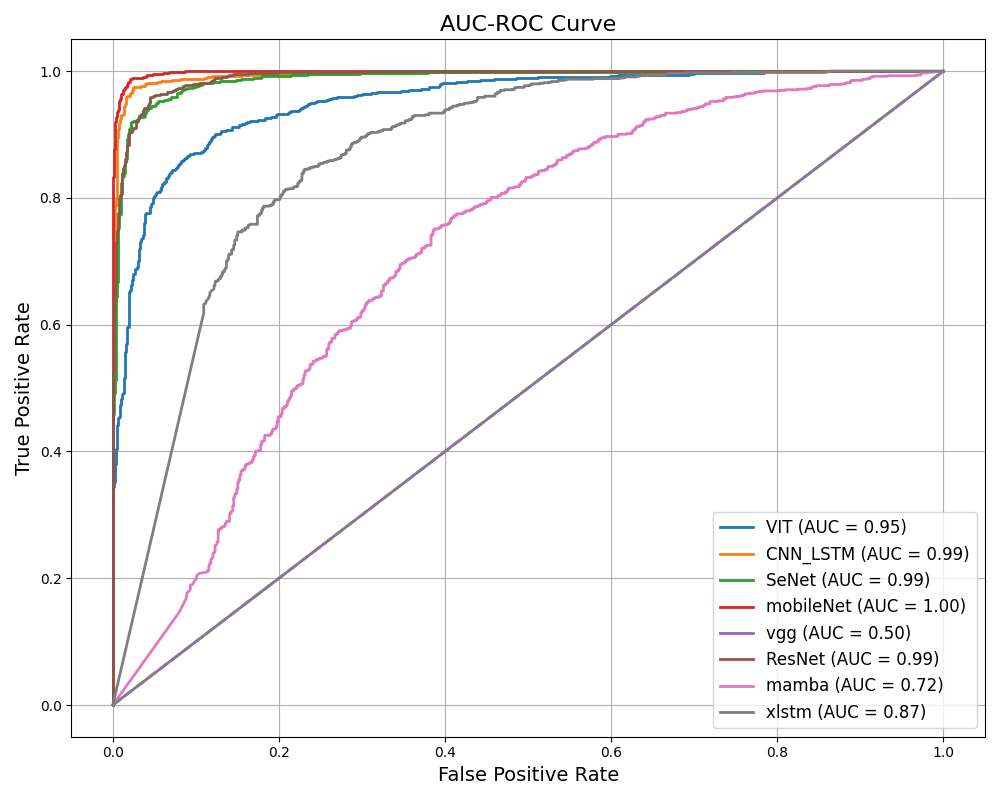}
    \caption{AUC-ROC graph for different models evaluated on FakeMusicCaps.}
    \label{fig:auc-roc}
\end{figure}

We conduct the systematic evaluation with the following classification models reviewed in Section II - Related work: QSVM, ResNet18, VGG, SeNet, MobileNet, CNN with LSTM, and ViT. The hyperparameters for all the models are shown in Table \ref{tab:hyperparameters}. 
\begin{table}[t!]
\centering
\caption{Hyperparameter settings. Different models are using different learning rate for best performance with hyperparamter finetuning.}
\label{tab:hyperparameters}
\begin{tabular}{l|c}
\hline
\textbf{Hyperparameter} & \textbf{Value} \\ \hline
Batch Size             & 64              \\
Epochs       & 10             \\
GPU & Tesla V100 \\
input size & 224 $\times$ 224 \\
Learning Rate          & 1e-3, 1e-4, 1e-5          \\
optimiser & Adam \\\hline
\end{tabular}
\end{table}
For the all deep learning architectures, we adopt the default values specified in their original papers. For QSVM, we set the kernel to `poly,' the degree to 2, and the coefficient $c_0$ to 1.
All performances are shown in ``Table \ref{tab:model_performance}'', and an AUC-ROC graph is shown in ``Figure \ref{fig:auc-roc}''.

\begin{table}[h]
\centering

%\resizebox{\columnwidth}{!}{
\begin{tabular}{lrrr}
\hline
\textbf{Model}       & \textbf{Training Time (s)} & \textbf{Accuracy} & \textbf{F1} \\ \hline
QSVM\cite{singh2021detection}                  & 649                        & .914              & .924              \\
ResNet18\cite{he2016deep}             & 96                         & .924              & .924              \\
VGG \cite{simonyan2014very}                 & 268                        & .902             & .902              \\
SeNet \cite{hu2018squeeze}             & 88                         & .930              & .930              \\
MobileNet \cite{howard2017mobilenets}           & 99                         & \textbf{.968}              & \textbf{.968}              \\
CNN+LSTM  \cite{hochreiter1997long}           & 83                         & .916              & .917              \\
ViT   \cite{dosovitskiy2020image}               & 518                        & .886              & .886              \\ 
Mamba   \cite{gu2023mamba}              &        22                  & .690              & .682              \\ 
xLSTM   \cite{beck2024xlstmextendedlongshortterm}               &       594                  &      \underline{.810}        &   \underline{.810}            \\ \hline
\end{tabular}%}
\caption{Training time, Accuracy, and F1 systematic evaluation comparison of models on the FakeMusicCaps dataset}
\label{tab:model_performance}
\end{table}

The presented results indicate generally strong performance across the models evaluated. MobileNet demonstrates superior efficacy, achieving an accuracy and F1 measure of 0.968 while maintaining a relatively brief training time of 99 seconds. This highlights its capability as a computationally efficient, yet good performance model for this task. In contrast, the xLSTM model demonstrates markedly lower performance. While Mamba exhibits marginally improved performance, its results remain suboptimal, which may due to the simplicity of its architectural leading to its shorter training time. This discrepancy suggests that the VGG architecture is not well-suited to the dataset or the specific task, although it performs adequately in other image-related tasks. The suboptimal results may also indicate that xLSTM combination of attention mechanism and long-short term mechanism prevent convergence to an optimal solution. Meanwhile, ResNet18, SeNet, CNN+LSTM, and VGG exhibit comparable and relatively strong performance, with ViT demonstrating slightly inferior results, potentially due to the overhead introduced by its attention mechanism. One plausible explanation is that ViT processes audio spectrograms in patches and relies on global attention, which may make it less sensitive to fine-grained local spectral patterns that are critical in some audio samples. In contrast, ResNet’s convolutional layers excel at capturing local correlations and short-term structures based on this empirical results. While we do not claim a definitive causal mechanism, this patch-based global attention characteristic of ViT may contribute to its subset-dependent performance. QSVM, while yielding moderate performance, is hindered by high computational complexity due to its quadratic kernel.

\subsubsection*{Out-of-domain test}Though the performance seems strong, we test the trained models on M6, both their generalised out-of-domain test subset (f) and the other remaining subsets (a-e), with the results shown in Table\ref{tab:outofdomain}.

\begin{table}[t!]
\centering
\caption{Out-of-domain model performance on M6. (f) means subset (f) in M6 while (o) means the other remaining ones.}
\label{tab:outofdomain}
% \resizebox{\columnwidth}{!}{
\begin{tabular}{lrrrr}
\hline
\textbf{Model} & \textbf{Acc(f)} & \textbf{F1 (f)} & \textbf{Acc(o)} & \textbf{F1 (o)} \\
\hline
QSVM\cite{singh2021detection} & .671 & \textbf{.758} & .780 & .626 \\
ResNet18\cite{he2016deep}  & \textbf{.685} & .693 & \textbf{.781} & \textbf{.778} \\
VGG\cite{simonyan2014very} & .667 & .533 & \underline{.274} & \underline{.118} \\
SeNet\cite{hu2018squeeze}  & .513 & .526 & .687 & .776 \\
MobileNet\cite{howard2017mobilenets} & .429 & .356 & .772 & .732 \\
CNN+LSTM \cite{hochreiter1997long}& \underline{.392} & \underline{.316} & .776 & .737 \\
ViT \cite{dosovitskiy2020image}  & .623 & .630 & .762 & .754 \\
Mamba   \cite{gu2023mamba}      &.623    &   .625                         & .656              & .670              \\ 
XLSTM   \cite{beck2024xlstmextendedlongshortterm}               &     .405        &      .422       &   .710            &  .712             \\ \hline
\end{tabular}
% }
\end{table}

The results indicate that all models experience a performance drop, highlighting both the poor generalisation capabilities of the systematic evaluation and the inherent challenges of the task. Specifically, in the out-of-domain test, MobileNet exhibits inferior performance compared to ResNet18. This discrepancy could be attributed to MobileNet's lightweight architecture, which may reduce its ability to capture complex features. This observation underscores ResNet18 performs the best. Overall, as a trade-off between in-domain and out-of-domain performance, we find that ResNet18 achieves the best balance for systematic evaluation: it performs comparably to MobileNet in in-domain settings while exhibiting substantially stronger generalisation in out-of-domain scenarios. Also, it emphasises the need for more structured guidance to enable models to learn truly distinguishable features.

Additionally, performance across the two subsets reveals that some models perform better on subset (f), while others excel on subsets (a-e). This inconsistency suggests that models may be capturing random or spurious features rather than intrinsic ones, highlighting the need for more robust models that can learn fundamental, task-specific representations. Future research should prioritize the design of models capable of effective generalization and focus on extracting intrinsic features. Furthermore, XAI could play a crucial role in guiding the identification of these intrinsic features.

\subsubsection*{Multimodal model}
We also test multimodal models on music with the lyrics subset (bc) of M6. We select an audio processor Wav2Vec 2.0 \cite{baevski2020wav2vec}, the default one on Huggingface and a text processor mBERT, a multilingual version of BERT \cite{devlin2018bert} to process the music. The extracted features are concatenated and fed into a Multi-Layer Perceptron (MLP) \cite{hinton1986mlp}, with optimisation occurring solely during MLP training. Results are presented in Table \ref{tab:multimodality}.

\begin{table}[htbp]
\centering

\begin{tabular}{lrr}
\hline
\textbf{Model} & \textbf{accuracy} & \textbf{F1}   \\
\hline
SVM & .917 & .915 \\
ResNet18 & .883 & .883 \\
ViT & .667 & .630 \\
Multimodal & .975 & .975 \\
\hline
\end{tabular}
\caption{Performance comparison of multimodal models with others on M6. The other three performances are provided in \cite{li2024m6multigeneratormultidomainmultilingual}.}
\label{tab:multimodality}

\end{table}

\begin{figure}
    \centering
    \includegraphics[width=0.9\linewidth]{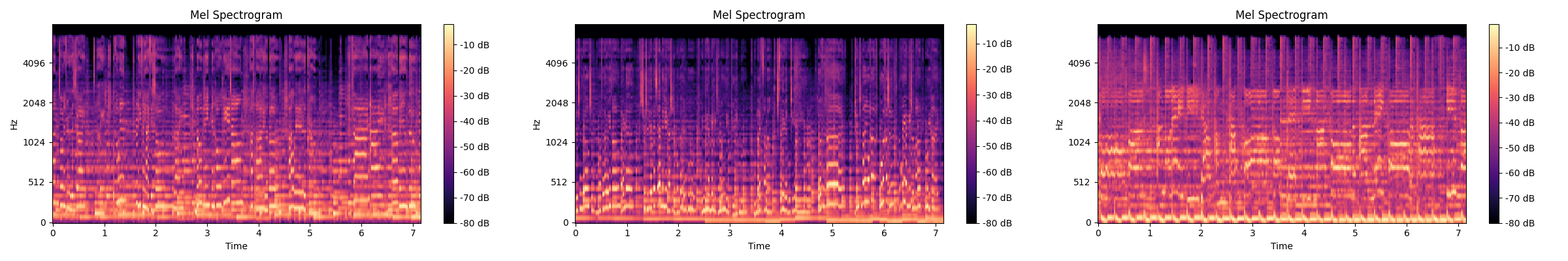}
    \caption{Mel-spectrograms of songs from English, Chinese, and Japanese cultures (left to right).}
    \label{fig:multiculture}
\end{figure}

The results show improved performance with multimodal models, indicating that relying solely on melody may be insufficient. Incorporating additional modalities, such as lyrics, enhances model performance. Additionally, we present mel-spectrograms of songs from different cultures, specifically Chinese, English, and Japanese in Figure \ref{fig:multiculture}. The comparison shows no substantial perceptual differences across these cultures, suggesting that, from an audio perspective, the deepfake detection task does not exhibit explicit cultural bias. We argue that deepfake detection should instead focus on properties intrinsic to the music itself, which may vary widely in stylistic characteristics, some of which correlate with cultural musical traditions. We have performed case study below to check the results in details.

\subsubsection*{Case study}
We have selected one test case to analyse the results. Specifically, we focus on a challenging case where the probabilities produced by the models are near the decision boundary (test ID 241). Given the dataset split strategy described earlier, with a random state of 42, this case is presented in Table \ref{tab:case study}.

\begin{table}[ht]
    \centering
    % \resizebox{\columnwidth}{!}{
    \begin{tabular}{cccc}
    \hline
        Model & Predicted Probability & Predicted label & Ground truth label \\
        \hline
        CNN+LSTM & .848 & 1 & 1\\
        mamba & .991 & 1 & 1 \\
        mobileNet & .498 & 0 & 1 \\
        ResNet & .542 & 1 & 1 \\
        SeNet & .455 & 0 & 1 \\
        VIT & .892 & 1 & 1 \\
        xlstm & 1.00 & 1 & 1 \\
         \hline
    \end{tabular}
    % }
    \caption{Model performance for test case 241 with predicted probability comparison}
    \label{tab:case study}
\end{table}

The results suggest that this test case is more recognisable when sequence-based models are used. Models incorporating sequential information, such as CNN+LSTM, Mamba, and XLSTM, achieve higher predicted probabilities, correctly classifying the instance as positive. In contrast, architectures that rely primarily on spatial features, such as MobileNet and SeNet, struggle to capture the underlying patterns, leading to lower confidence scores or incorrect classifications. This indicates that temporal dependencies play a crucial role in accurately identifying this test case. However, there are additional cases where frequency-based information plays a dominant role in classification, while temporal information is less critical. This highlights the need for future research to explore XAI techniques for multimodal models, enabling a deeper understanding of how different modalities contribute to the decision-making process.

\section*{Explainable AI}
\label{sec:xai}

\begin{figure*}[t]
  \centering
  \includegraphics[width=1\textwidth]{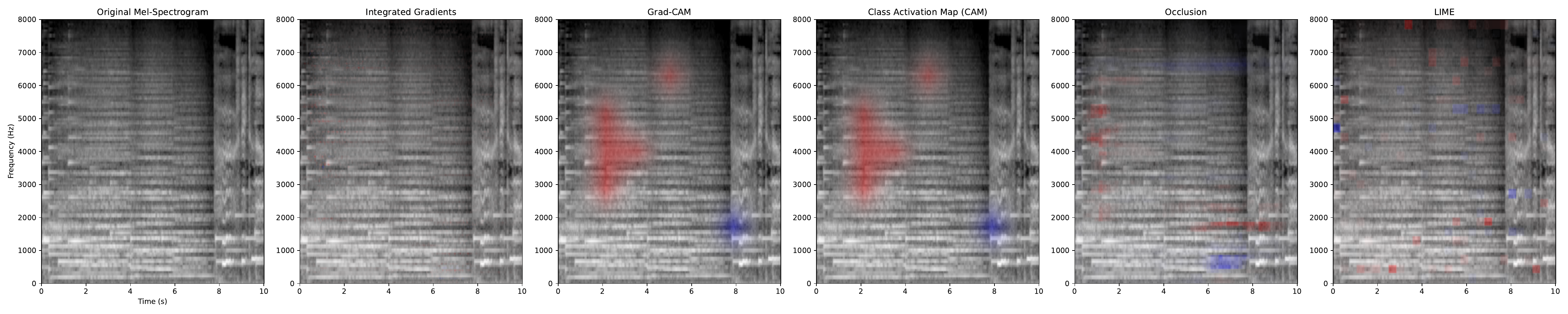}
  \caption{Visualisation of an audio sample using different XAI techniques on the ResNet18 model.}
  \label{fig:xai_heatmap}
\end{figure*}

\begin{figure*}[t]
  \centering
  \includegraphics[width=1\textwidth]{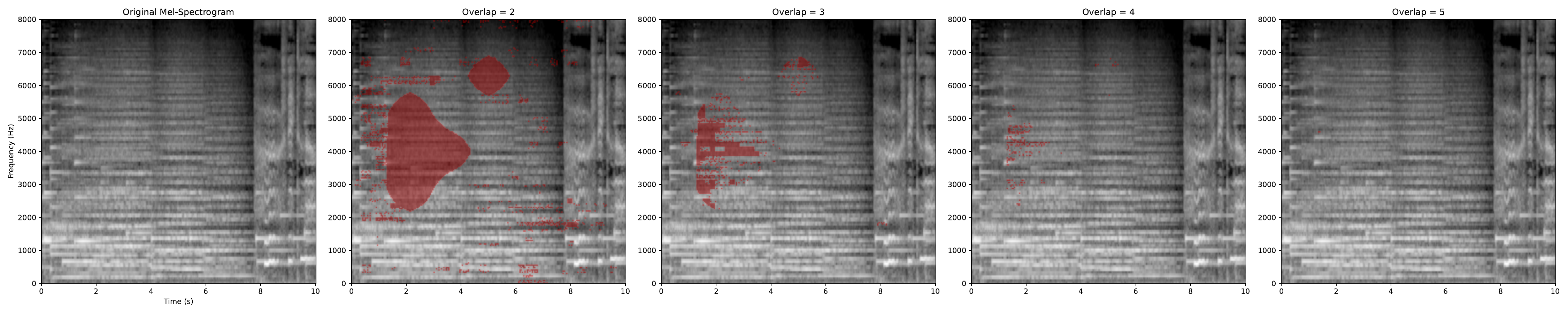}
  \caption{Visualisation of overlapping high-contribution regions using multiple XAI techniques on the ResNet18 model}
    \label{fig:xai_overlap_heatmap}
\end{figure*}

% We also explore the explainability of the model in the task of detecting deepfake music. We select ResNet as the representative model for the audio modality due to its strong performance and compatibility with various XAI techniques. We adopt several widely used explainability methods, including IG, LIME, CAM, Grad-CAM, and Occlusion. These methods highlight the regions of input features that the model considers important analyse the decision-making process. We conduct both qualitative and quantitative analyses. \textbf{Notably, we introduce a novel fidelity experiment in the quantitative analysis by combining multiple XAI techniques.} This aims to evaluate the effectiveness and reliability of these explainability techniques in this task through heatmap observations and performance validation based on feature modification.

We also explore the explainability of the model in the task of detecting deepfake music. We select ResNet as the representative model for the audio modality due to its strong performance and compatibility with various XAI techniques. We adopt several widely used explainability methods, including IG, LIME, CAM, Grad-CAM, and Occlusion. While the aforementioned methods are standard in vision tasks, applying them individually to audio spectrograms often yields high variance due to the continuous and high-dimensional nature of frequency features. To address this, we extend these atomic methods into a consensus-based framework. Instead of relying on a single explanation, we compute the intersection of top-scored regions across varying explanation mechanisms. This strategy filters out method-specific noise and identifies 'robust features' that are consistently deemed critical, thereby elevating XAI from simple application to a more reliable methodological workflow for music detection. 

Based on this ensemble XAI strategy, we conduct both qualitative and quantitative analyses. Notably, to validate this strategy, we perform a fidelity assessment by combining the overlapping regions from these techniques. This aims to evaluate the effectiveness and reliability of our proposed consensus approach through heatmap observations and performance validation based on feature modification.

\subsection*{Qualitative Analysis}
In the qualitative analysis, we investigate the patterns and consistency of different XAI techniques by examining the explainability heatmaps of a single sample. Specifically, we selected a representative deepfake audio sample from the validation set where the model demonstrated high confidence (softmax probability \(>\)0.9) in its prediction. %To ensure reproducibility, we provide the file path of the selected sample and the exact steps in the supplementary material. 
Using the trained model weights, we generated predictions for this sample and applied various XAI techniques to produce heatmaps, as shown in Figure \ref{fig:xai_heatmap}. 
%BS: Do not "randomly" do things in papers w/o providing seed numbers or a link to assure others can REPRODCUE THIS PROPERLY!
% qs: I changed. We will provide the path of the "random" sample on code. And I also calculate the confidence which is over 0.96.
The heatmaps highlight high-contribution regions critical to the model’s classification. A red-blue transparent colour map is used to visualise the magnitude and polarity of contributions. Red shows positive contributions, while blue represents negative ones. Thresholding is applied to some XAI techniques, focusing on the top 10\% of positive and negative contributions. Other regions are made transparent to avoid over-coverage and enhance clarity.

Through observation of the heatmaps, we identify the following patterns. First, the high-contribution regions generated by different XAI techniques show a degree of consistency. For example, in the 0-4 second mid-frequency range and the 6-8 second low-frequency range, IG, CAM, Grad-CAM, and Occlusion all highlight similar high-contribution regions. This suggests that these regions are likely key to the model's classification decisions. Additionally, the 6-8 second low-frequency range is highlighted as a blue negative contribution region, consistent with the characteristics of the selected audio sample, a real sample correctly predicted as the positive class. In this case, the negative contribution regions in the heatmap represent features that the model identifies as interfering with the classification of the sample as `real music'. In other words, the model detects certain uneven frequency distributions and categorises these features as negative contributions correctly. The filtering of potentially disruptive features strengthens the model's confidence in recognising the sample as real. This behaviour indicates that the model can enhance its recognition of positive samples by effectively ignoring anomalous regions. However, from a perceptual perspective, there is a noticeable break at the 8-second mark in this sample. This might result from a musical variation or pause. The observation that most XAI models consider this segment negatively correlated with the positive class suggests a potential limitation: the model may interpret frequent musical breaks as spectral anomalies rather than inherent structural elements. This implies that the current model's decision logic might be misaligned with the essence of musicality in regards to rhythmic pauses.% in both positive and negative samples.

Additionally, different XAI techniques show notable differences in their focus on feature breadth and distribution. Grad-CAM and CAM typically concentrate on localised regions of the audio signal. This is likely due to their reliance on the final feature maps of CNNs, which often capture higher-level local information. In contrast, LIME and IG provide finer-grained explanations by focusing on broader regions, covering the entire frequency spectrum. These differences may result from the model’s internal mechanisms and parameter choices. For instance, the baseline used in IG considerably impacts the calculation of feature contributions. In this study, the baseline is set to a zero tensor, which may amplify changes between the input and the baseline. This approach likely guides the model to prioritise regions with greater variations. LIME, however, generates random feature masks to evaluate their influence on the model’s output. While this method enhances input coverage, it introduces uncertainty, particularly in the interpretation of local details, where the stochastic nature of mask generation may compromise the consistency of explanations.
% The qualitative analysis helps to understand the model’s attention patterns intuitively. It also provides an important foundation for the design of subsequent quantitative experiments. In the next section, we use these XAI techniques to conduct single and superposition fidelity experiments.

While our qualitative analysis sheds light on the model's attention patterns, we acknowledge the limitations of post-hoc methods, as prior work has shown that saliency maps may remain unchanged under model randomisation and that audio explanations often fail to reflect true decision cues \cite{adebayo2018sanity, praher2021veracity}. Despite these concerns, we chose these methods. First, they remain widely adopted and allow comparisons with existing literature. Second, they offer complementary perspectives: while CAM-based methods focus on final-layer activations, LIME and IG can reveal more fine-grained patterns. Third, by triangulating across multiple techniques, we reduce reliance on any single explanation and enhance robustness.

Nevertheless, we emphasise that these techniques mainly indicate where the model attends in the input space, not why. That is, they reveal spatial saliency on the Mel-spectrogram, but not the underlying semantic attributes such as rhythm regularity, pitch stability, or timbral consistency that may have influenced the decision. Bridging this gap between attribution and semantic meaning remains an open challenge in XAI for audio.

\subsection*{Quantitative Analysis}
To verify the reliability of the XAI results, we design quantitative experiments based on fidelity. The core idea is to mask the regions deemed most important by the model and then reassess its classification performance. This approach quantifies the actual contribution of these regions.% to the model’s decision-making on deepfake detection task.

In the single-technique fidelity experiments, we mask the top 10\% of high-contribution regions identified by each XAI method, replacing them with zero-filled values. This tests the impact of removing critical regions. Masking is applied to over a thousand validation samples, and metrics such as accuracy, F1, and recall are re-evaluated. As shown in Table \ref{tab:xai_fidelity1}, all XAI techniques experience a significant drop in performance. For IG, prediction accuracy falls to 50.8\%.

\begin{table}[htbp]
    \centering
    
    \begin{tabular}{l c c c }
        \hline
        \textbf{Visualisation} & \textbf{Accuracy} & \textbf{F1} & \textbf{Recall}  \\ \hline
        Raw Spectrogram & .907 & .920 & .962 \\
        IG & .508 & .641& .508 \\
        Grad-CAM & .805 & .851 &.987 \\ 
        CAM & .830 & .870 &.966\\ 
        Occlusion & .590 & .732& .790 \\ 
        LIME & .566 &.698 &.883 \\ \hline
    \end{tabular}
    \caption{Fidelity experiments of  masking high-contribution regions from single XAI technique}
    \label{tab:xai_fidelity1}
\end{table}
To quantitatively validate the effectiveness of our proposed ensemble XAI strategy, we conduct fidelity experiments on the consensus regions. This involves combining the heatmaps generated by 2 to 5 XAI techniques and masking only the overlapping regions. i.\,e., when the combination size is 2, we mask areas covered by at least two XAI techniques. When the combination size is 3, we mask areas covered by at least three techniques, and so on. 

Figure \ref{fig:xai_overlap_heatmap} shows the visualisation of overlapping high-contribution regions on one audio sample. These regions are identified by multiple XAI techniques on the ResNet18 model. The first sub-figure shows the original Mel-spectrogram of the sample. The next four sub-figures highlight the overlapping regions identified by at least 2, 3, 4, and 5 XAI techniques. As the combination number increases, the overlapping regions become smaller. the highlighted areas concentrate on a small, specific region around 1.5 seconds and 4.5 kHz.

We also evaluate the overall impact of masking these overlapping regions. We calculate several metrics using five independent experiments. The results are shown in Table \ref{tab:xai_fidelity2}.  The "Avg Mask" column shows the average percentage of masked regions. This value drops as the combination size increases. For instance, the average masked area decreases from 29.6\% at size 2 to only 0.43\% at size 5. This trend matches the pattern observed in Figure \ref{fig:xai_overlap_heatmap}. The accuracy shows the model's performance on the validation set. Average accuracy improves from 48.6\% to 80.0\% as the masked area reduces.  The p-value quantifies the statistical significance of accuracy changes between successive combination sizes. The improvement between sizes 3 and 4 is significant ($p = 0.004$), as is the change between sizes 4 and 5 ($p = 0.01$). In contrast, the change between sizes 2 and 3 shows less statistical significance ($p = 0.15$).  To illustrate the trade-off further, the mask reduction shows the relative decrease in masked area compared to the previous combination size. This reduction reaches 87.4\% at size 5. The accuracy change  highlights the relative improvement in accuracy. The most notable improvement occurs between sizes 3 and 4, with an increase of 36.7\%.

These results demonstrate a clear trade-off between reducing the masked area and maintaining accuracy stability. As the combination size increases, the masked area decreases substantially, from 29.6\% to 0.43\%. However, accuracy improves gradually and stabilises. This outcome confirms that the Ensemble Consensus Strategy effectively isolates the high-fidelity signal from the noise present in individual explainers. It achieves a balance between minimising the masked area and maintaining stable performance.

% We calculate the average percentage of masked regions and evaluate performance metrics for each combination size on the validation set, as shown in Table \ref{tab:xai_fidelity2}. The results indicate that combining overlapping regions from different XAI techniques considerably
% %BS: p-value?! Test method?! Then, do not write "significantly" :) I changed... 

% %QS:Bjoern I'm not sure whether p value is necessary here as I want to emphsis the trade-off of mask size and accuracy metric. But I added it on p-value. Although the p = 0.15 during 2 to 3 size. As the random of some XAI techniques. So the accuracy is not quite stable. But it gets better after size 3.

% reduces the masked area compared to using individual techniques. When the combination size reaches four, the masked area accounts for only 3\% of the image. For combinations of five techniques, the masked area drops to just 0.43\%. However, performance does not decline too much as the masked area decreases. Specifically, it exceeds the results of masking based on single XAI techniques in some cases. This demonstrates that our approach effectively identifies the most critical regions. Additionally, it highlights a degree of consistency among different XAI techniques in recognising high-contribution areas.

% qs: acc show in .486 ± .044? need suggestion
\begin{table}[htbp]
    \centering
    
    % \resizebox{\linewidth}{!}{%
    \begin{tabular}{c c c c c c }
        \hline
        \textbf{Combination Size} & \textbf{Avg Mask}  & \textbf{Accuracy}  &\textbf{p-value} & \textbf{Mask Reduction} & \textbf{Accuracy Change} \\ \hline
        2 & 29.6$\pm$7.9 & 48.6$\pm$4.4 & -     & -        & -     \\
        3 & 10.2$\pm$2.2 & 55.6$\pm$5.2 & 0.15    & 65.5     & 14.4  \\
        4 & 3.42$\pm$0.4 & 76.0$\pm$3.0 & 0.004     & 66.5     & 36.7  \\
        5 & 0.43$\pm$0.3 & 80.0$\pm$2.2 & 0.01  & 87.4     & 5.3   \\
        \hline
    \end{tabular}%
    % }
    \caption{Fidelity experiments of combined overlapping regions from multiple XAI techniques}
    \label{tab:xai_fidelity2}
\end{table}

\section*{Discussion}
\label{sec:discussion}
This study establishes a comprehensive systematic evaluation for the task of MGMD. A diverse array of architectures are evaluated ranging from traditional CNNs to emerging State Space Models. Our extensive quantitative experiments confirm that while binary classification is achievable with high accuracy using deep neural networks within specific domains. By systematically assessing these models, we validate the feasibility of deploying automated detection systems whilst highlighting the performance degradation observed when models confront unseen generative algorithms or hybrid datasets. Furthermore, the integration of multimodal features and XAI in our systematic evaluation. Beyond binary classification, this study has remarkable implications for real-world deployment, such as automated copyright management and content tagging on streaming platforms. A robust detector can protect human artists by filtering massive volumes of AI-generated uploads.

Based on the systematic evaluation results, several practical observations can be drawn to guide the future research. Regarding model choice, CNN-based architectures should be favored when training data and representations are fixed, as they consistently achieve strong performance across both in-domain and out-of-domain evaluations, whereas Transformer-style models (such as ViT) do not show a clear advantage and in some cases underperform despite higher computational cost. Concerning input representations, all models are evaluated using the same spectrogram configuration, and the strong performance achieved by multiple CNN variants indicates that standard spectrogram settings are already sufficient to support competitive MGMD performance, without requiring task-specific feature redesign. However, in the multimodal setting, a simple early-fusion strategy based on concatenating frozen audio and text embeddings yields consistent improvements over unimodal baselines, suggesting that lightweight fusion mechanisms constitute an effective and practical starting point, while more complex multimodal modeling can be explored in future work.

From a musicological perspective, our XAI analysis points to a probable divergence between model behaviour and human auditory perception. As observed in the specific case study (Fig. \ref{fig:xai_heatmap}), the model appeared to penalise a 'musical break' (silence), potentially treating it as a spectral defect. In contrast, human listeners perceive such pauses as structural rhythmic elements. Although establishing this pattern as a systematic bias requires larger-scale expert annotation, our findings provide a compelling hypothesis: that current models heavily rely on low-level spectral artefacts rather than high-level musical semantics. Future MGM detection should aim to validate and bridge this gap by incorporating music-theoretic features such as beat tracking and structural segmentation to align model judgements closer to human aesthetic perception.

\begin{figure}
    \centering
    \includegraphics[width=0.85\linewidth]{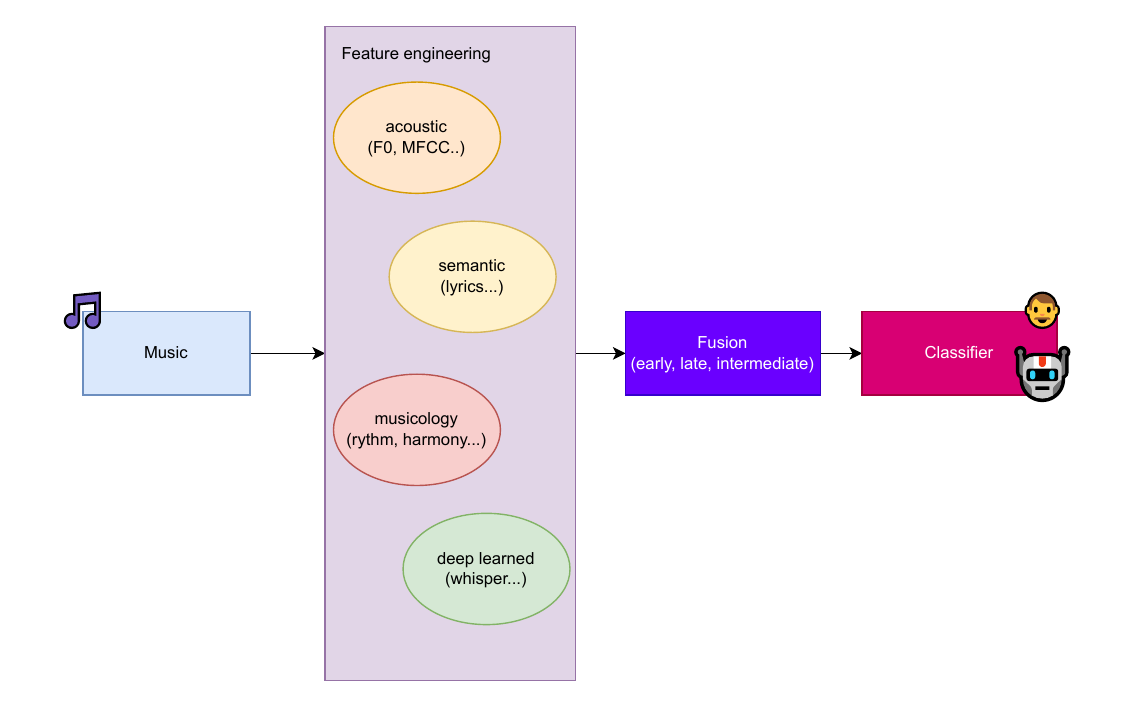}
    \caption{Proposed MGMD pipeline. The music is first processed to extract diverse features, including acoustic, semantic, musicological, and deep learning–based representations. These features are subsequently processed and fused, and the fused representations are ultimately used for classification.}
    \label{fig:pipeline}
\end{figure}

Note that all analyses based on the evaluation results are conducted within a general-purpose audio processing framework, and we adopt a standard binary supervised learning setting commonly used in audio deepfake detection. The results indicate that a substantial degree of shared knowledge exists between audio and music deepfake detection. Therefore, as this work aims to provide an early systematic evaluation, we do not introduce novel methodologies, loss functions, or model architectures. Nevertheless, the XAI techniques reveal important regions relevant to the MGMD task, and, together with the musicological analysis, they suggest a promising direction for future research, as illustrated in Figure \ref{fig:pipeline}. Specifically, acoustic features emphasising particular frequency regions, raw deep learning-based audio representations, semantic information (e.g. derived from lyrics), and musicological features such as rhythm could be carefully extracted and more effectively fused using specially designed mechanisms, potentially inspired by cross-attention. These fused representations could then be further processed in a post-feature stage for classification.

Also, we acknowledge several limitations.  First, We relied on Mel-spectrograms and basic early fusion for multimodal tasks. While sufficient for establishing baselines, as it is currently the most widely adopted representation in modern audio models, including popular systems such as Whisper \cite{radford2022robustspeechrecognitionlargescale} and recent audio LLMs \cite{peng2024survey}, this approach may overlook temporal structural features that are crucial in musicology. We focus on FakeMusicCaps and M6, as they are among the few publicly available music deepfake datasets, whereas other existing datasets are not publicly accessible or have legal constraints. Moreover, inviting professional musicians to judge based on musicology could greatly add more insight to future feature engineering. In addition, as noted in our qualitative analysis, current XAI methods primarily visualise spatial attention rather than providing semantic reasoning. They highlight spectral regions but do not explicitly explain decision logic based on high-level musical attributes like rhythm or harmony.

\section*{Conclusion}
\label{sec:conclusion}
In this work, we systematic evaluate the performance of ten models spanning traditional machine learning, deep neural networks, and Transformer-based methods on the MGMD task, providing a robust set of baselines for future research. Additionally, we explored multimodal approaches, demonstrating their potential in advancing music feature analysis by incorporating complementary modalities such as lyrics.% alongside audio features.

To assess model robustness and generalisability, we conducted out-of-domain testing and utilised XAI techniques to analyse whether the models learnt intrinsic features. Furthermore, we proposed a novel ensemble XAI strategy to interpret model decisions effectively. Our findings reveal that existing models often fail to leverage fundamental musicological features, such as stops and structural elements, indicating a gap in their ability to capture domain-specific knowledge. This highlights the need for future research to integrate domain-specific knowledge into model architectures, paving the way for more interpretable and robust AI systems in music analysis. Additionally, MGMD based on alternative music representations, such as symbolic or MIDI-based formats, could be explored in future work once more suitable datasets become available.

% \section*{Acknowledgements (not compulsory)}

% Acknowledgements should be brief, and should not include thanks to anonymous referees and editors, or effusive comments. Grant or contribution numbers may be acknowledged.
\section*{Data Availability}
All data generated or analysed during this study are included in FakeMusicCaps: https://paperswithcode.com/dataset/musiccaps, and M6 \cite{li2024m6multigeneratormultidomainmultilingual} published article.
\section*{Funding}
This research was partially supported and funded by the Munich Center for Machine Learning and the Munich Data Science Institute.
\section*{Author contributions statement}

Yupei Li: lead the project, write manuscripts, do experiments;
Qiyang Sun: write manuscripts, do experiments;
Hanqian Li: do experiments;
Lucia Specia: supervision, check manuscripts, provide GPUs;
Bj\"orn W. Schuller: major supervision, re-write manuscripts, funding

\bibliography{sample}

\end{document}